\documentclass{aastex631}

\received{June 1, 2020}
\revised{\today}
\accepted{\today}
\submitjournal{ApJ}

\shorttitle{Lyman continuum}
\shortauthors{Yuan et al.}

\graphicspath{{./}{figures/}}

\def\hbeta{H$\beta$}

\def\h{~$h^{-1}$}
\def\m{~$\mu$m}
\def\R25{$R_{25}$}

\def\CIII {\ion{C}{3}]}
\def\CIV  {\ion{C}{4}}

\def\HeII {\ion{He}{2}}
\def\NV   {\ion{N}{5}}

\def\OII  {[\ion{O}{2}]}
\def\OIII {[\ion{O}{3}]}

\def\Ionone {{\it Ion1}}
\def\Iontwo {{\it Ion2}}
\def\Hubble{{\it Hubble}}

\begin{document}

\title{CDFS-6664: A \textbf{candidate} of Lyman-continuum Emission at $z \approx 3.8$ detected by HDUV}

\correspondingauthor{Fang-Ting Yuan}
\email{yuanft@shao.ac.cn}

\author{Fang-Ting Yuan}
\affiliation{Key Laboratory for Research in Galaxies and Cosmology, Shanghai Astronomical Observatory, Chinese Academy of Sciences, 80 Nandan Road, Shanghai 200030, China}

\author{Zhen-Ya Zheng}
\affiliation{Key Laboratory for Research in Galaxies and Cosmology, Shanghai Astronomical Observatory, Chinese Academy of Sciences, 80 Nandan Road, Shanghai 200030, China}

\author{Ruqiu Lin}
\affiliation{Key Laboratory for Research in Galaxies and Cosmology, Shanghai Astronomical Observatory, Chinese Academy of Sciences, 80 Nandan Road, Shanghai 200030, China}
\affiliation{University of Chinese Academy of Sciences, No. 19A Yuquan
Road, Beijing 100049, China}

\author{Shuairu Zhu}
\affiliation{Key Laboratory for Research in Galaxies and Cosmology, Shanghai Astronomical Observatory, Chinese Academy of Sciences, 80 Nandan Road, Shanghai 200030, China}
\affiliation{University of Chinese Academy of Sciences, No. 19A Yuquan
Road, Beijing 100049, China}

\author{P. T. Rahna}
\affiliation{Key Laboratory for Research in Galaxies and Cosmology, Shanghai Astronomical Observatory, Chinese Academy of Sciences, 80 Nandan Road, Shanghai 200030, China}

\begin{abstract}

{We report the detection of Lyman Continuum (LyC) emission from the galaxy, CDFS-6664, at $z=3.797$ in a sample of Lyman break galaxies with detected [OIII] emission lines. The LyC emission is detected with a significance $\sim 5\sigma$ in the F336W band of the {\Hubble} Deep UV Legacy Survey, corresponding to the $650-770${\AA} rest-frame. The light centroid of the LyC emission is offset from the galaxy center by about $0.2$\arcsec{} ($1.4$ pkpc). The {\Hubble} deep images at longer wavelengths show that the emission is unlikely provided by low-redshift interlopers. The photometric and spectroscopic data show that the possible contribution of an active galactic nucleus is quite low. Fitting the spectral energy distribution of this source to stellar population synthesis models, we find that the galaxy is young ($\sim50\mathrm{Myr}$) and actively forming stars with a rate of $52.1\pm4.9~M_{\odot}\mathrm{yr^{-1}}$. The significant star formation and the spatially offset LyC emission support a scenario where the ionizing photons escape from the low-density cavities in the ISM excavated by massive young stars. From the nebular model, we estimate the escape fraction of LyC photons to be $38\pm7$\% and the corresponding IGM transmission to be $60\%$, which deviates more than $3\sigma$ from the average transmission. The unusually high IGM transmission of LyC photons in CDFS-6664 may be related to a foreground type-2 quasar, CDF-202, at $z=3.7$, with a projected separation of 1.2\arcmin\ only. The quasar may have photoevaporated optically thick absorbers and enhance the transmission on the sightline of CDFS-6664. }

\end{abstract}

\keywords{galaxies:high-redshift - galaxies:evolution - reionization}

\section{Introduction} 
\label{sec:intro}

During the epoch of reionization, the neutral and opaque intergalactic medium (IGM) transforms to ionized and transparent. The process requires Lyman continuum (LyC, $h\nu>1\mathrm{Ryd}$) photons that can ionize the neutral hydrogen gas. Currently, the nature and properties of the sources of the reionization is still in debate. The two sources generally accepted to keep the IGM ionized are quasars and star-forming galaxies, but the relative contribution of these sources and its dependence on the cosmic time are still not clear \citep[e.g.,][]{madau2014,dayal2020}. Previous works find that the number density of quasars may decline rapidly at $z>3$ \citep[e.g.,][]{hopkins2007}. Therefore, star-forming galaxies may be the main source of the LyC photons at high redshift \citep[e.g.,][]{steidel2001}.
 
The ability of LyC photons from galaxies to ionize the IGM depends on how many of the ionizing photons can escape from the galaxy environment considering the absorption of the interstellar gas and dust, that is, the escape fraction of the LyC photons ($f_\mathrm{esc}$). However, there is still a lack of knowledge about the average escape fraction from the star-forming galaxy population \citep[e.g.,][]{grazian2017}. 

The most straightforward method to measure $f_\mathrm{esc}$ is to detect LyC photons from galaxies directly.  
In recent years several works have detected LyC sources from nearby universe to redshift$\sim 4$ \citep[e.g.,][]{debarros2016,shapley2016,vanzella2016,bian2017,izotov2018,fletcher2019,nakajima2020, saha2020}. However, the success rate of searches of LyC emitters at high redshift is still quite low. The detection of distant LyC sources is difficult in the following aspects: (1) Contamination from foreground sources can produce false LyC detections; (2) LyC visibility is quite stochastic, because of the stochastic character of the IGM transmission, the irregular geometrical distribution of gas, dust, and stars in galaxies and the short $f_\mathrm{esc}$ duty circle over cosmic time; (3) Current facilities allows only to probe LyC from relatively luminous galaxies, which intrinsically has lower escape fraction of photons \citep{inoue_iwata2008,mostardi2015,siana2015,vanzella2016}.  As a result, there are only a few LyC emitters that have been detected at $z>3$, although these galaxies are crucial for us to understand the mechanism that allows ionizing photons to escape. 

As the accumulation of observations, the chance to detect LyC sources also increases. Deep and large surveys enable us to find more LyC sources. Researches also find that galaxies with certain features are more likely to be LyC emitters. 
Ly$\alpha$ lines, \OIII/\OII{} ratios, and ultraviolet (UV) absorption lines are useful probes to select LyC emitter candidates \citep[e.g.,][]{Heckman2011,jaskot2014,nakajima_ouchi2014,verhamme2015,alexandroff2015,dayal2018,nakajima2020}. These pre-selection methods further increase the detection rate of LyC emitters.

In this work, we report a detection of LyC emission arising from a emission-line galaxy at $z=3.797$ based on HST observations. The photometric and spectroscopic data of this object enable us to make a detailed analysis on its properties.  
Throughout this paper, we use pkpc and
pMpc (ckpc and cMpc) to indicate the proper (comoving) distances. The AB magnitude system
$\mathrm{AB}=-2.5\log (f/\mathrm{Jy})+8.9)$ and a cosmology of $\Omega_\mathrm{tot}$, $\Omega_\mathrm{M}$, $\Omega_{\Lambda}= 1.0, 0.3, 0.7$ with $H_{0} = 70 \mathrm{km~s^{-1}~Mpc^{-1}}$ are used.

\section{Data}

\begin{figure*}
    \centering
    \includegraphics[width=0.9\textwidth]{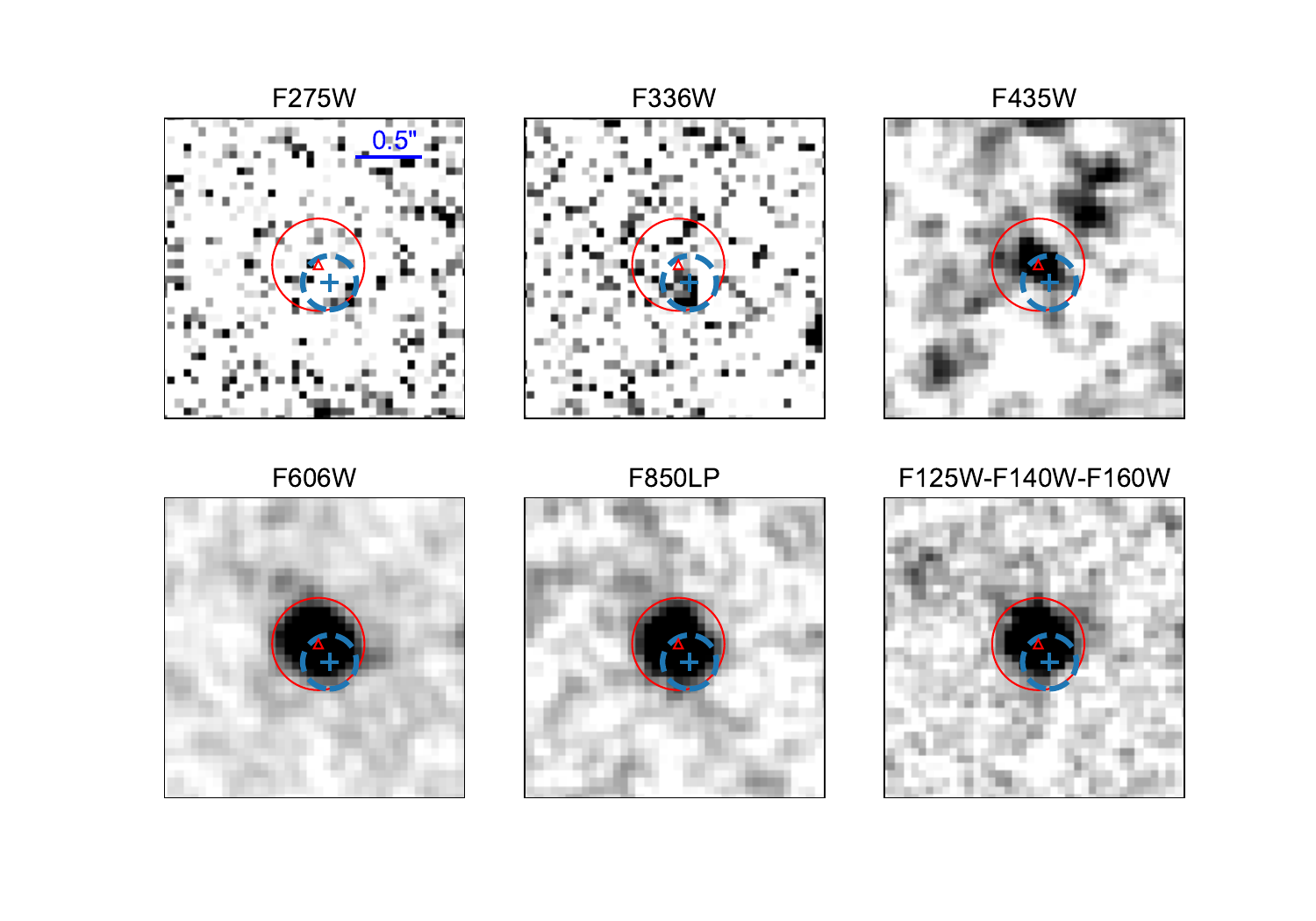}
    \caption{HST multiwavelength images of CDFS-6664 with a box size of 2.4\arcsec$\times$2.4\arcsec. The center of the F336W-band detection is marked by the red plus. The $0.4\arcsec$ aperture of the source detected by F336W-band is indicated in each image with the blue dashed line. The red triangle on each panel indicates the center of the source measured from the F160W image. The red circle shows an aperture of 0.7\arcsec diameter, corresponding to a physical scale of $\sim 5.0$ kpc at $z=3.797$.}
    \label{fig:cdfs6664stamp}
\end{figure*}

The LyC emitter candidate, named CDFS-6664, was selected from an emission-line Lyman break galaxy (LBG) sample at $z\sim 3.5$, which includes a compilation of spectroscopic observations using near-infrared instruments on large ground telescopes (VLT SINFONI and Keck I MOSFIRE). A more detailed description of the sample can be found in our previous work \citep{yuan2019}. 

We use the publicly available images from the Hubble Deep UV Legacy Survey \citep[HDUV,][]{oesch2018} at F275W and F336W bands with $5\sigma$ depths of 27.6 and 28.0 mag, respectively, in the GOODS-S field. In addition to the rest-frame UV data, we also use the 3D-HST images from F435W to F160W \citep{grogin2011,koekemoer2011,skelton2014} to analyze the escape fraction of the LyC photons. All these images are registered to the WFC3 mosaics based on their header WCS and re-binned to a 0.06\arcsec/pix scale.
{
We show the images of CDFS-6664 from F275W to F160W bands in Figure \ref{fig:cdfs6664stamp}. Visually, the source can be marginally seen in the F336W band at the lower right position relative to its F435W, F606W, and F160W detection. There is no confusion source nearby to contaminate the emission. }

{The source can be detected in the F336W band by SExtractor \citep{ba1996} with the same parameters used in the CANDELS `hot' mode \citep{guo2013}. The position of the F336W detection is shown in Figure \ref{fig:cdfs6664stamp}. The signal-to-noise ratio (S/N) is $7.9$ for the isophotal measurement, 6.5 in an aperture of 0.4\arcsec, and 5.0 in an aperture of 0.7\arcsec{}. CDFS-6664 has an aperture magnitude of $27.9\pm0.2$ in the $0.7\arcsec$ aperture. We also run Sextractor in dual-image mode using an aperture of $0.7\arcsec$. The detection image is a noise-equalized combination of the F125W, F140W,
and F160W images. Following the method used by \citet{skelton2014}, we find that the total magnitude is $27.7\pm0.2$ after applying the aperture to total correction by a factor of $f_\mathrm{F160W,total}/f_\mathrm{F160W, aper}$. This total magnitude is used in the SED analysis in Section \ref{sec:analysis}.
}

{In fact,} CDFS-6664 has {already been included} in the HDUV and Hubble Legacy Fields (HLF) catalogs \citep{oesch2018,whitaker2019}. {The total magnitude at the F336W band} is $27.7\pm0.2$ mag from the HDUV catalog and $27.5\pm0.2$ mag from HLF. Our measurement is closer to that given by the HDUV catalog. In the following, we take the {total} magnitude of CDFS-6664 in the F336W band as $27.7\pm0.2$ mag. This value corresponds to a flux of $3.0\pm0.6~\times10^{-31}$erg s$^{-1}$ cm$^{-2}$ Hz$^{-1}$.

\begin{figure*}
    \centering
    \includegraphics[width=\linewidth]{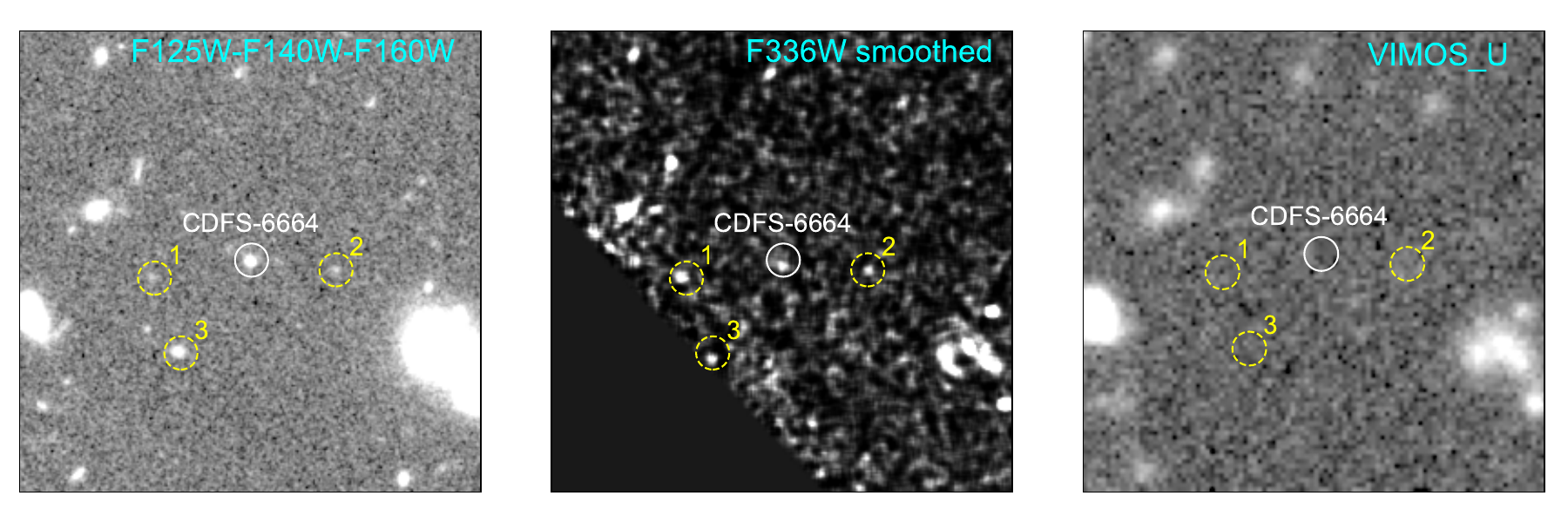}
    \caption{Galaxies that have been detected in the F336W band but not in the VIMOS $U_V$ band. Left: The 3D-HST source detection image, which is a noise-equalized combination of the F125W, F140W, and F160W images. Middle: F336W image that has been smoothed with a Gaussian kernel with $\sigma=0.3\arcsec$. Right: VIMOS $U_V$ image. The box size is $20\arcsec\times20\arcsec$. }
    \label{fig:uvimos_det}
\end{figure*}

{We note that CDFS-6664 has not been detected by VIMOS $U_V$-band \citep{nonino2009}, whose depth is about 28.1 mag. According to the catalog data, we find that the completeness of the VIMOS $U_V$ to detect the $5\sigma$ F336W sources is about $70\%$. The VIMOS $U_V$-band probes wavelength slightly redder than the F336W band, but still well within the Lyman continuum domain at $z\approx3.8$. A galaxy with similar redshift and magnitude, named as \Ionone{}, is detected with S/N$\sim 6.6$ in the $U_V$-band \citep{vanzella2015}. The fact that there is no such detection for CDFS-6664 poses doubts about the reliability of the F336W detection. However, the VIMOS $U_V$ coverage within the CDFS field is not uniform. CDFS-6664 is coincidentally located in a gap region where the total exposure time is only $<$80\% of that for \Ionone{} (66933 s versus 84209 s) and about a half of the maximum\footnote{According to the single-exposure images provided by ESO Science Archive (\url{http://archive.eso.org/cms.html}).}. Moreover, this galaxy is surrounded by several bright galaxies that elevate the level of the background, causing difficulty to detect faint sources. We show in Figure \ref{fig:uvimos_det} that there are about four galaxies (including CDFS-6664) that have been missed by the VIMOS $U_V$ image in this region. }

\section{Spectroscopic and Photometric Properties}
\label{sec:analysis}
\begin{figure*}
    \centering
    \includegraphics[width=0.7\linewidth]{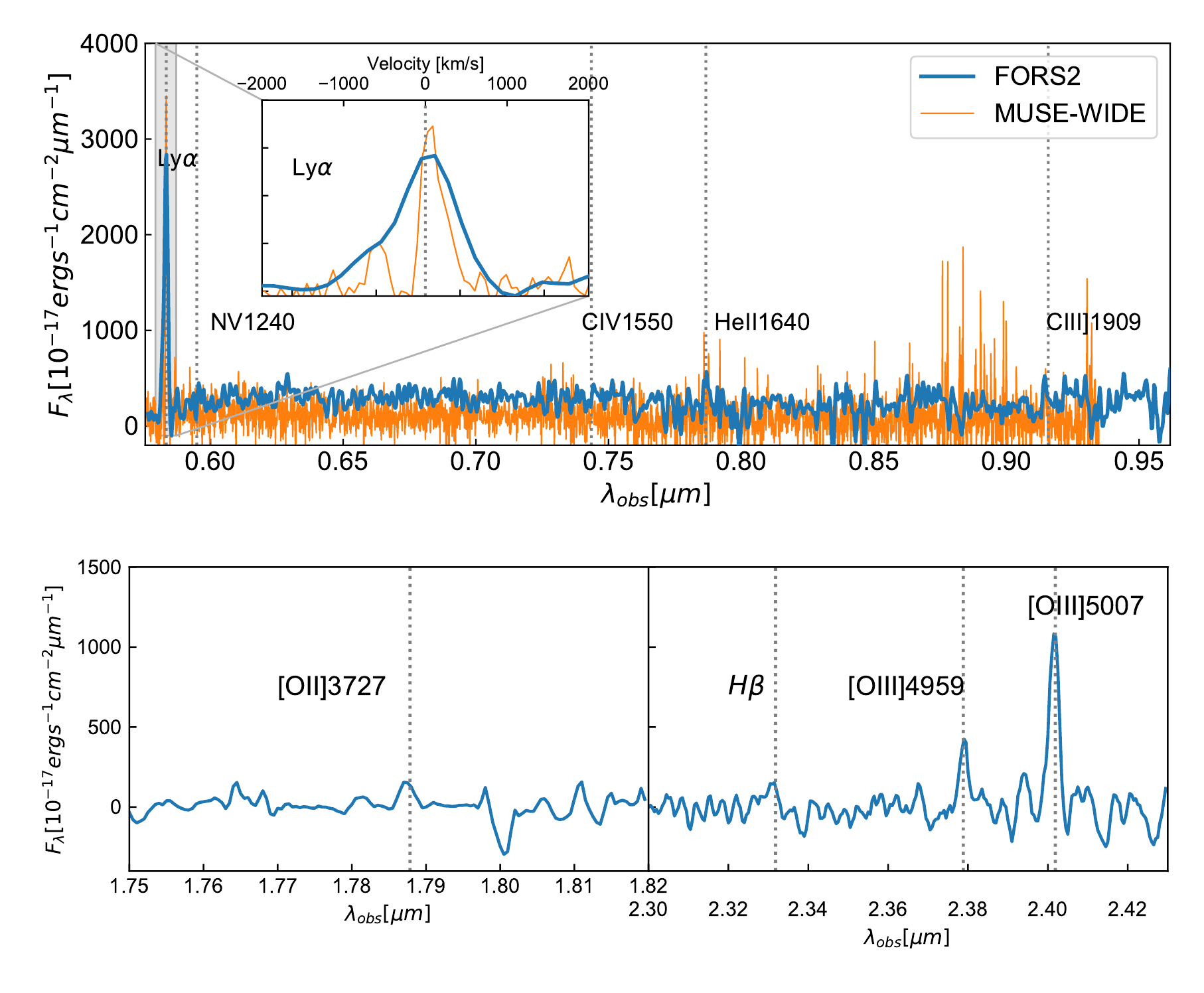}
    \caption{Upper panel: Optical spectrum of CDFS-6664 from VLT/FORS2 observation by \citet{vanzella2006} (thick blue line) and from MUSE-Wide observation by \citet{urrutia2019} (thin orange line). Lower panel: Near infrared spectrum (smoothed) of CDFS-6664 from VLT/SINFONI observation by the AMAZE project \citep{maiolino2008}. {In both panels, the vertical dotted line marks the observed wavelength of each emission line based on the systemic redshift ($z=3.797$, calculated from \OIII).}}
    \label{fig:spec}
\end{figure*}

The optical spectrum by VLT/FORS2 observation \citep{vanzella2006} shows that this galaxy is also a Ly$\alpha$ emitter, with Ly$\alpha$ equivalent width $\sim 148\pm37${\AA} (Figure \ref{fig:spec}). The spectral resolution of VLT/FORS2 ($R\sim660$) is not high enough for us to carry out further analysis on the structure of the Ly$\alpha$ line. {Fortunately, CDFS-6664 has also been observed by the MUSE-Wide survey \citep{urrutia2019}. The spectrum is shallower than the VLT/FORS2 one but is with a higher spectral resolution ($R\sim3000$). The Ly$\alpha$ line shows double peaks. The blue and red peaks are offset from the systemic velocity by $\sim-600$ km/s and $100$ km/s, respectively. The Ly$\alpha$ emission emerging at the systemic velocity is an important tracer for LyC leakers because it indicates clumpy, multi-phase systems with non-unity covering fractions \citep{naidu2021}. For CDFS-6664, the central escape fraction of Ly$\alpha$ emission within $\pm 100$ km/s of the systemic velocity $f_\mathrm{cen}$ is about $28$\%, consistent with the LyC leakers with $f_\mathrm{esc}>20\%$ \citep[e.g.,][]{vanzella2020,naidu2021}.}


High-ionization emission lines like \NV~1240 and \CIV~1550 have not been detected at more than the 3$\sigma$ level, while the \HeII~1640 and \CIII~1909 lines are detected. {Based on the FORS2 spectrum, the \HeII~1640 and \CIII~1909 are estimated as $5.9\pm1.4\times 10^{-18}$ and $5.4\pm1.4\times 10^{-18}$ erg s$^{-1}$cm$^{-2}$, respectively.
From the ratio of \CIII~1909/\HeII~1640 ($\sim0.9\pm0.3$), and following
\citet{feltre2016},} this source is classified as a star-forming galaxy lying in the same region occupied by low metallicity galaxies of \citet{stark2014}.

CDFS-6664 has also been observed in the near-infrared band using VLT/SINFONI by the 
AMAZE project \citep{maiolino2008}. The near-infrared spectrum shows that this galaxy has 
\OIII/\OII $\sim 10$ (Figure \ref{fig:spec}). 
The large Oxygen ratio and the strong \OIII{}4959,5007 emission lines are both fit the profile predicted by the photon ionization models for an LyC candidate emitter \citep{jaskot2014,nakajima_ouchi2014}.

We then analyze the panchromatic SED of CDFS-6664 from UV to mid-infrared bands using the broadband data from HST and \textit{Spitzer} observations. We also include the flux of \OII, \hbeta, \OIII4959, and \OIII5007 in the fitting to provide more constraints on the stellar mass and SFR \citep{yuan2019}. {In the above analysis, we use the total magnitude in the SED fitting to keep the measurements of the continuum and the emission lines consistent.}

\begin{figure*}
    \centering
    \includegraphics[width=0.95\linewidth]{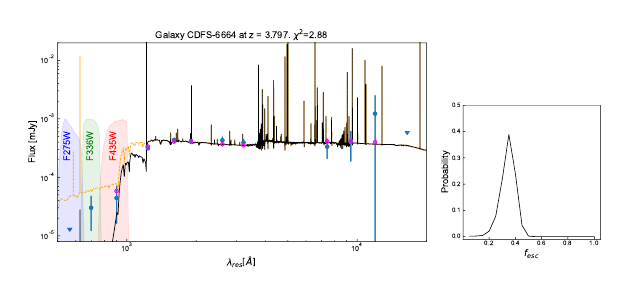}\\
    \caption{SED fitting for CDFS-6664. Left panel: The blue dots are the observed broadband fluxes. The down triangles are the observed upper limits. The black solid line is the spectrum of the best-fit model. The magenta circles are the broadband fluxes estimated from the best-fit model. The spectrum with no IGM absorption is shown as the orange dashed line. The filter response curves of F275W, F336W, and F435W are shown as blue, green, and red shaded areas, respectively. Right panel: The probability density distribution of $f_\mathrm{esc}$.}
    \label{fig:sed}
\end{figure*}

We use Code Investigating GALaxy Emission \citep[CIGALE,][]{burgarella2005,noll2009,boquien2019} to fit the SED with models of stars, gas, and dust. The SED of CDFS-6664 can be fitted well assuming a delayed star formation history (Figure \ref{fig:sed}). The most significant discrepancy between the data and model comes from the UV bands because the average IGM model \citep{meiksin2006} used in CIGALE is not consistent with the observation. We discuss the IGM effect in Section \ref{sec:discussion}. 

The results of SED fitting show that CDFS-6664 is a low mass galaxy with $\log ({M_{*}/M_{\odot})}=9.15\pm0.15$ and star formation rate of $52.1\pm4.9~M_{\odot}\mathrm{yr^{-1}}$. The stellar population of this galaxy is quite young, with an average age of $0.05\pm0.01$ Gyr. The dust attenuation for this galaxy is quite low, with $E(B-V)_s=0.10\pm0.05$. 

CIGALE also obtains an escape fraction based on the SED. The fitting result gives $f_\mathrm{esc}=0.38\pm0.07$. Examining the shape of the probability density function (Figure \ref{fig:sed}), we find the parameter $f_\mathrm{esc}$ is well constrained in the fitting. The degeneracy between $f_\mathrm{esc}$ and IGM transmission is broken by the constraints introduced from the emission line data, assuming
\begin{equation}
   L_{\mathrm{H}\beta}(f_\mathrm{esc})=L_{\mathrm{H}\beta}(0)\times\frac{1-f_\mathrm{esc}}{1+0.6f_\mathrm{esc}}.  
\end{equation}
where $L_{\mathrm{H}\beta}$ is the luminosity of the \hbeta{} line emission \citep{inoue2011}. The electron temperature $T_e$ is assumed to be $10^4~$K \citep{inoue2011,boquien2019}. The ionization parameter $U$ is assumed to have $\log U=-2.5$, corresponding to $\log q_\mathrm{ion}/\mathrm{cm\,s}^{-1} = 8.0$, consistent with the value for local LyC leakers \citep{nakajima_ouchi2014}. 

\section{Discussion}
\label{sec:discussion}

\subsection{LyC morphology}

The spatially resolved images of HST enable us to analyze the morphology of CDFS-6664. Using GALFIT to fit the image of the F160W band, we found that the profile of the galaxy can be well described by a single Sersic component, with a reduced $\chi^2$ of $1.038$. The Sersic index of the galaxy is 0.83, suggesting a disk-like morphology. CDFS-6664 has a  physical half-light radius of 1.3 kpc, which is more extended than previously found LyC emitters at $z>3$, such as \Ionone\ \citep[$z=3.795$]{vanzella2015,ji2020}, \Iontwo\ \citep[$z=3.2$]{debarros2016,vanzella2016}, J0121+0025 \citep[$z=3.24$]{ marques-chaves2021}.

Comparing the F336W image with other bands, we find that the LyC emitting region is confined to a small area that deviates from the center of the galaxy. The images of CDFS-6664 of other bands show no sign of any possible low-redshift interloper at this position. {Similar offsets have been found in \Ionone, which is at a redshift very close to CDFS-6664, and several other high redshift galaxies \citep[e.g.,][]{micheva2017}.} The deviation from the center of the LyC emission may imply a clumpy distribution of the ISM caused by the strong star-forming activities in this galaxy.

The deviated LyC emission also implies that the normal stacking method may underestimate the escape fraction of the LyC photons because the stacking is always done by matching the centers of different galaxies. If LyC photons are not from the center, the signals cannot be enhanced by such stacking.

\subsection{The source of LyC photons}

We analyze the photometric and spectroscopic data of CDFS-6664 to investigate the possibility of an AGN contribution. We summarize the following aspects suggesting that the source is more likely to be a stellar origin:     

First, CDFS-6664 is in the sample of the AMAZE project \citep{maiolino2008,troncoso2014}, in which LBGs that host AGNs are discarded based on the UV/optical spectra, X-ray data or the MIPS 24$\mu$m fluxes. Furthermore, as mentioned in Section \ref{sec:analysis}, the line ratio of \CIII~1909/\HeII~1640 ($\sim0.6$) shows that this source locates in the same region occupied by star-forming galaxies. Considering the facts that the UV/optical spectra cannot remove type-2 AGN, and that X-ray/MIPS 24$\mu$m are both redshift-dependent so that they may miss a low-luminosity AGN, these pieces of evidence cannot rule out the possibility of an AGN-origin for the LyC photons. Nevertheless, these data show that the probability of AGN contributing to the LyC photons for this source is quite low.

Second, the SED fitting shows that  {the UV spectral slope $\beta$ is $-2.25\pm0.03$}. This slope is not extremely steep but indicates a rather blue UV color. Therefore, it is likely that the LyC photons are generated from the massive young stars in UV bright clumps.  

Finally, from the morphological fittings of the profiles of F160W and F435W images using GALFIT, we find that the morphology of CDFS-6664 is more close to disk-like. The off-central morphology in the F336W band resembles the extended UV emission found in the star-forming regions of the nearby galaxies, supporting the scenario that the LyC photons are coming from the star-forming regions. The disk-like and clumpy morphology further reduces the possibility that it is an AGN dominant source.

\subsection{LyC escape fraction and IGM transmission}

The escape fraction of a galaxy can be calculated by the following equation:

\begin{equation}
\label{equ:fesc}
    f_\mathrm{esc}=\frac{(L_\mathrm{1500}/L_\mathrm{800})_\mathrm{int}}{(L_\mathrm{1500}/L_\mathrm{800})_\mathrm{obs}}
    \times T_\mathrm{IGM,F336W}^{-1} \times 10^{-0.4\,A_\mathrm{1500}},
\end{equation}
where $L_\mathrm{1500}$ and $L_\mathrm{800}$ are the non-ionizing ($\lambda_\mathrm{rest}=1500$\AA) and ionizing ($\lambda_\mathrm{rest}=800$\AA) luminosity from the galaxy, respectively.  $T_\mathrm{IGM,F336W}^{-1}$ is the IGM transmission at the F336W band. $A_\mathrm{1500}$ is the attenuation at $1500$\AA. Here we assume that the dust affects only the non-ionizing UV (i.e.,
$A_{800}=0$) \citep[see e.g.,][]{steidel2018}. 

In previous works, by assuming the transmission of the IGM and the intrinsic ratio of the UV to LyC photons ($(L_\mathrm{1500}/L_\mathrm{800})_\mathrm{int}$), one can estimate the value of $f_\mathrm{esc}$. In this work, instead of using the assumption on the IGM transmission, we estimate the $f_\mathrm{esc}$ by assuming the ISM properties. As mentioned in Section \ref{sec:analysis}, the SED fitting provide constraints on the $f_\mathrm{esc}$ by the nebular radiative transfer model and obtain the value to be $\sim0.4$. The result is also consistent with the $f_\mathrm{esc}$ derived from the \OIII/\OII{} ratio using the relation given by \citet{nakajima_ouchi2014} or \citet{nakajima2020}. This is not surprising because both of the methods are based on the nebular models. 

Adopting the value $f_\mathrm{esc}$ as $40\%$, the $T_\mathrm{IGM}$ is estimated to be $\sim 60\%$ from Equation \ref{equ:fesc}. Even if we use the $99$th percentile value of the $f_\mathrm{esc}$ probability density distribution ($\sim45\%$), the $T_\mathrm{IGM}$ is about $\sim 53\%$, which is higher than the $99$th percentile of the simulated IGM transmission value (about $35\%$) at such a redshift based on the results of \citet{steidel2018}. 
As also shown in the SED fitting, the average IGM model given by \citet{meiksin2006} cannot fit the observed fluxes for F435W and F336W bands.    
The discrepancy can be explained by the extremely stochastic property of the IGM \citep[e.g.,][]{inoue_iwata2008}. Recent works also find that the mean IGM transmission may not suitable for LyC emitters \citep{bassett2021,prichard2021}.

\begin{figure}
    \centering
    \includegraphics[width=0.7\linewidth]{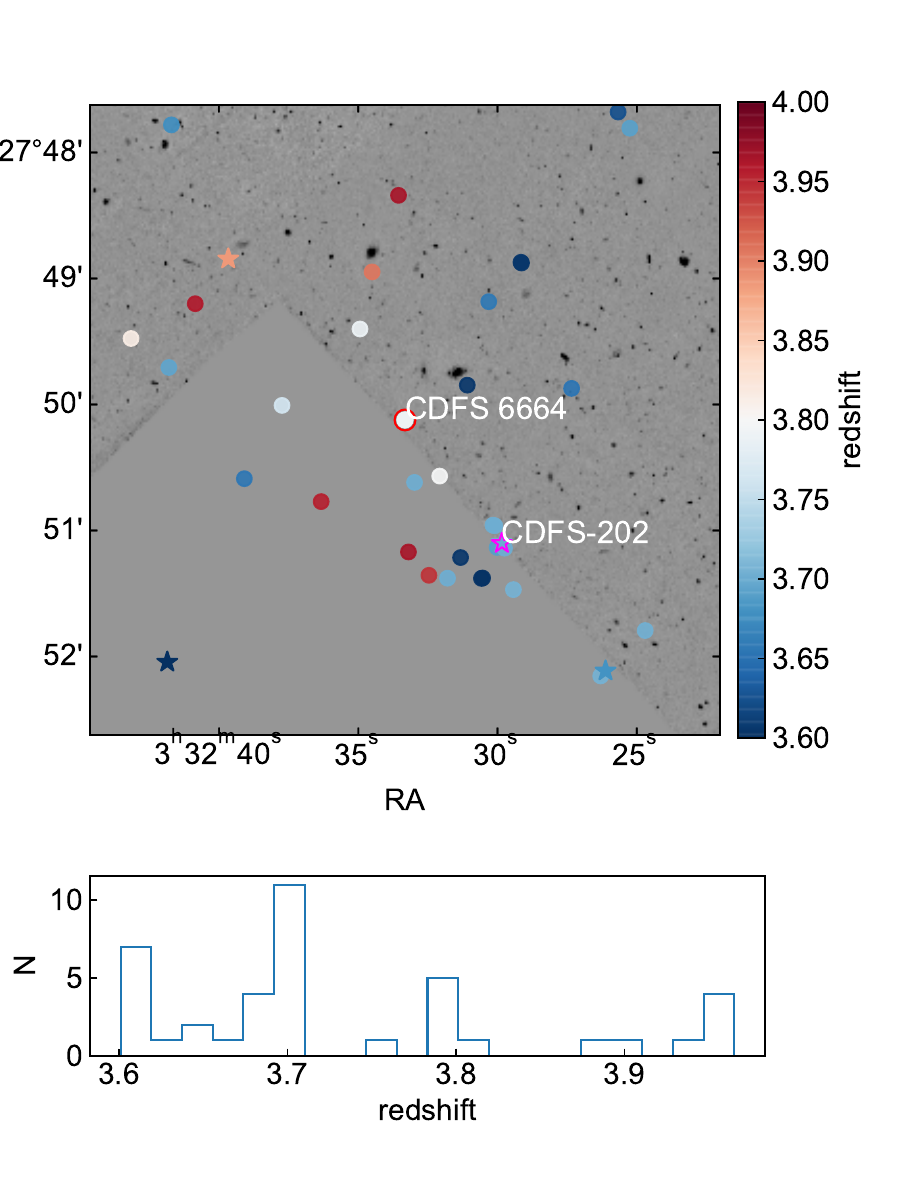}
    \caption{Upper panel: Spatial distribution of CDFS-6664 and CDFS-202 over-plotted on the F336W images ($5\arcmin\times 5\arcmin$). The colored dots and stars show the galaxies and quasars with spectroscopically confirmed redshifts $3.6\leq z \leq 4.0$, respectively. Lower panel: Redshift distribution of the galaxies shown in the upper panel. }
    \label{fig:distribution}
\end{figure}

In addition to the inhomogeneous nature of the IGM, we consider the possibility that a foreground quasar close to the sightline may ionize its surrounding IGM and open a path for the LyC photons of CDFS-6664 to the observers. We check the quasar catalog and find that the famous type-2 quasar, CDFS-202 \citep{norman2002}, is close to the sightline of CDFS-6664, with a project separation of $0.5$ pMpc ($1.2\arcmin$, Figure \ref{fig:distribution}). CDFS-202 is located at redshift $3.71$. The line-of-sight distance to CDFS-6664 is about 64.9 cMpc. 
Previous works show that luminous quasars can photoevaporate optically thick absorbers within a $\sim 1$ pMpc radius or so \citep[e.g.,][]{hennawi_prochaska2007}. Therefore, it is quite possible that the ionization bubble of CDFS-202 affect the IGM on the sightline of the CDFS-6664 and enhances the transmission of LyC photons. 

{We also plot the sources with known spectroscopic redshift higher than 3.6 near CDFS-6664. The redshift distribution shows a narrow peak at $z=3.7$, indicating a clustering of sources at this redshift, as shown in Figure \ref{fig:distribution}. } These sources may present extra ionizing background, further increase the transmission of the IGM at this region \citep[e.g.,][]{adelberger2003}.

\section{Summary}

This work reports the discovery of {a candidate} LyC emission from the LBG, CDFS-6664, at $z=3.797$. The galaxy has an \OIII/\OII{} value of about 10 and an asymmetric Ly$\alpha$ line profile, both indicating possible leakage of LyC photons. Based on the HDUV F336W band image, we found a detection at $27.9\pm0.2$ mag in an aperture of $0.7\arcsec$, corresponding to $650-770${\AA} rest-frame. summary

We further investigate the properties of this galaxy using the spectroscopic and photometric data. The data show the AGN contribution to the LyC emission is negligible. We derived from the SED fitting that the average age of the stellar populations is about $50$ Myr, and the SFR is $52.1\pm4.9~M_{\odot}\mathrm{yr^{-1}}$. These results show that the source of the LyC photons is more likely to be the massive young stars.

We break the degeneracy between the escape fraction and the IGM transmission by constraining the escape fraction using the nebular model and the observed \hbeta{} emission. In this way, the escape fraction is estimated to be $38\pm7\%$. Consequently, the IGM transmission for this galaxy is about $60\%$, exceeding the 99th percentile transmission ($35\%$) obtained by the simulation at this redshift. 

The unusually high transmission of the IGM along the line of sight of CDFS-6664 can be explained by the stochastic nature of the IGM. {Morever, the foreground ionizing sources may increase the transmission on the sight line. Specifically, we found that the quasar CDFS-202 is close enough to the sightline of CDFS-6664 ($0.5$ pMpc projected separation) to ionize the foreground IGM and increase the free path for the LyC photons. Whether LyC emitters and quasars are spatially correlated requires further investigation.}

Additionaly, the LyC emission of CDFS-6664 is offset from the galaxy center, indicating a clumpy distribution of the ISM in the galaxy. The spatially offset LyC emission implies that the escape fraction can be higher in some regions of galaxies than estimated from the entire galaxy. If the spatially offset LyC emission is normal in high-redshift galaxies, the commonly adopted stacking method by previous works may underestimate the escape fraction of LyC photons.

In addition to CDFS-6664, we have also detected a few suspicious off-center LyC signals at the F336W band for the emission-line galaxies in our sample. However, the current depth of the UV surveys ($\sim 28$ mag) is not able to confirm these detections. Further investigation of these sources requires deeper imaging in UV bands. In particular, the extreme deep field (XDF) project with the Multi-channel Imager (MCI) instrument carried by the Chinese Space Station Telescope (CSST) plan to reach the depth of $\gtrsim 29$ mag at its near-UV band \citep{yuan2021}. The data product will provide us an unprecedented chance to investigate the LyC emitters statistically.

\begin{acknowledgements}
We thank the anonymous reviewer for his/her helpful comments. This work is partly supported by the Funds for Key Programs of Shanghai Astronomical Observatory. FTY acknowledges support from the Natural Science Foundation of Shanghai (Project Number: 21ZR1474300). ZYZ acknowledges support by the National Science Foundation of China (11773051, 12022303), the China-Chile Joint Research Fund (CCJRF No. 1503 \& 1906) and the CAS Pioneer Hundred Talents Program. PTR acknowledges support by the CAS President's International Fellowship Initiative (PIFI) under the Grant No. E085201009.

This research has made use of the services of the ESO Science Archive Facility. Based on observations collected at the European Southern Observatory under ESO programme 168.A-0485. 

This work is based on observations taken by the 3D-HST Treasury Program (GO 12177 and 12328) with the NASA/ESA HST, which is operated by the Association of Universities for Research in Astronomy, Inc., under NASA contract NAS5-26555.

{This work is based on observations taken by the MUSE-Wide Survey as part of the MUSE Consortium.

Observations have been carried out using the Very Large Telescope at the ESO Paranal Observatory under
Program ID(s): 170.A-0788, 074.A-0709, and 275.A-5060.} 

\end{acknowledgements}

\bibliography{lyc_ref}{}
\bibliographystyle{aasjournal}

\end{document}